# GoAutoBash: Golang-based Multi-Thread Automatic Pull-Execute Framework with GitHub Webhooks And Queuing Strategy

Hao Bai

Department of Computer Engineering

University of Illinois at Urbana-Champaign, Champaign, IL, USA, 61820,

Haob2@illinois.edu

## ABSTRACT

Recently, more and more server tasks are done using full automation, including grading tasks for students in the college courses, integrating tasks for programmers in big projects and server-based transactions, and visualization tasks for researchers in a data-dense topic. Using automation on servers provides a great possibility for reducing the burden on manual tasks. Although server tools like CI/CD for continuous integration and *Hexo* for automated blog deployment have been developed, they're highly dedicated to certain functionalities and thus lack general usage. In this paper, we introduce a Golang-based automation framework that reacts to the events happening on GitHub in a multi-thread approach. This framework utilizes a queue to arrange the tasks submitted and execute each task with a thread in a preemptive manner. We then use the project GoAutoGrader to illustrate a specific implementation of this framework and its value in implementing high-freedom server applications. As Golang is developing in a rapid way because of its incredible parallel programming efficiency and a super-easy way to learn on the basis of C-like programming languages, we decide to develop this system in Golang.

## 1. Introduction

Automation is widely used to deal with repetitive and constructive tasks in the engineering world. In the pattern of agile development [1], programmers are required to use the Continuous Delivery/Continuous Integration approach (CI/CP) to shorten their development cycle and improve integration frequency. According to Fowler [2], CI is a software development practice where members of a team integrate their work frequently, usually each person integrates at least daily - leading to multiple integrations per day. Each integration is verified by an automated build (including test) to detect integration errors as quickly as possible. In web hosting integration, frameworks like *Hexo* for blog deployment is pretty popular among programmers for their simplicity of deployment and functionality to write a blog on a laptop and use one command to synchronize the content onto the server [3].

For the two cases mentioned above, the service provider can directly communicate with the client to finish the transaction. However, there is a special scenario where programmers push their code to a VCS (Version Control System) like GitHub and at the same time want their code to be examined using the version just pushed. In other words, the code on VCS should be the one to examine. In this case, we utilize the functionality of VCS called Webhook [4]. Most VCS providers give Webhooks, which raises an event to all the server domains specified when a certain action (like pushing, pulling, starring) is executed. From the view of the server which is to examine the code, it receives a signal from VCS and then triggers a series of scripts to accomplish its task.

Thus, we introduce our framework that executes a series of tasks after the client has triggered an event that raises a Webhook in VCS. In this paper, we use GitHub as the example of VCS and the push event as the Webhook trigger event. A graphical illustration of the scenario is shown in Figure 1. When the client pushes his repo to GitHub, a Webhook will be sent from GitHub to all the servers specified, including the one shown in the figure. The figure must be listening on a port using a REST API POST and should trigger a callback handler whenever it receives a POST request from the Internet. The callback handler then calls the executing function ("series of commands" in the graph) and execute the testing tasks, including pulling the latest repo from GitHub, doing regression tests on the code, running through the compilers, and generating reports, deploying the source code to the server, etc. When the scripts are done, the Go running environment generates a signal of ending the process and the server then pushes all the files generated onto the GitHub repo, which ends the workflow. Now that the new repo is accessible for common users of GitHub, they're free to access the resources simply by pulling the code down.

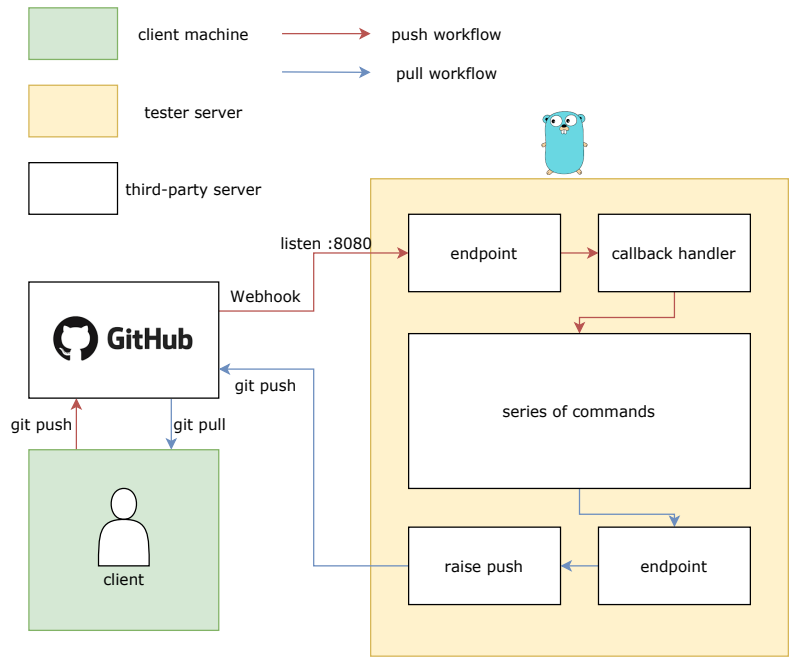

Figure 1. The general workflow of our framework. Three parties are concerned in our framework: client, VCS, and the tester server. In our example, the VCS is GitHub and we use Golang to implement the server.

Before we start explaining the details (and the parallelism patterns) in the implementation, there are some prerequisites to know about in Section 2. These concepts are crucial for understanding the code implemented in the project.

## 2. Related Works

In this part, we discuss concepts related to the core concepts of our system, including the Webhook system and Gin library in Golang, and the producer-consumer model used in our parallel programming task.

**2.1 The Webhook System**

Webhook is a signal sent to a specified server domain whenever an event specified is triggered. According to Handoyo [5], Webhook can work if there is some trigger based on post request or get request on the tester server, so after creating the subscription on VCS we still need to implement a listening service running on the tester server, waiting to receive Webhook signals from VCS.

Generally speaking, there are two approaches for the tester server to detect a change on VCS (more generally, another server): polling and Webhook. Polling strategies are various, but all of them are tester-server-driven, so it has to run all the time using another process or thread to check whether there are any content modifications on the VCS side. On the other side, the Webhook approach is event-driven or VCS-driven, so the server only needs to listen to a specific port and execute when a new trigger comes. This event-driven idiosyncrasy greatly ameliorates the performance of the server and makes the whole system more robust, as listening takes much fewer resources than polling.

Just like the overall framework, the Webhook system also concerns the three parties: tester server, VCS, and the client. A graphical illustration is shown in Figure 2.

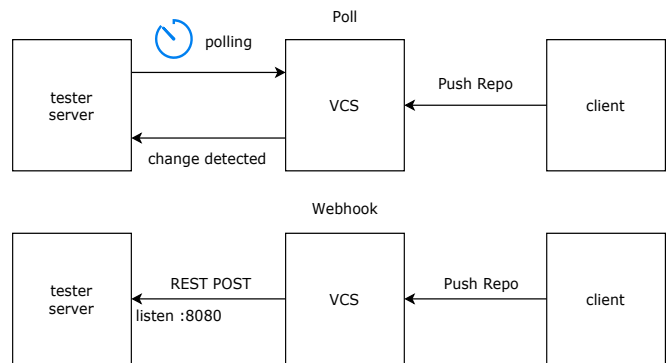

Figure 2. The difference between the polling strategy and the Webhook approach. The VCS in both methods takes client input randomly (just like in the real world), while the server need to respond for each new submission onto the VCS.

## 2.2 The Go Programming Language

Since its release on November 10, 2009, Golang has become one of the most popular programming languages in the Web programming field. Different from other server-side languages like JavaScript, it provides an incredible performance yet simpler grammar. The native http libraries include almost all of the required functionality for all the web tasks, from which we can see the design of the Go language is web-oriented. In our work, we use an even more powerful Web tool library called Gin, which will be discussed in Section 2.3.

Specifically, the most important reason why we choose to use Golang is its high performance for parallelism. Compared to other high-performance languages like C and node.js, Golang provides an inherent way for parallelism called GoRoutine. As our project requires a huge amount of parallelism to generate reports for different student requests, we choose Golang as our first choice. The built-in complement type Channel is also a great time-saver to implement a queue that auto-detects whether there is a new task and decides which thread (or consumer, as we use a producer-consumer model) should take this task.

## 2.3 The Gin Request Library

Representational State Transfer (REST) is now one of the most popular WebAPIs, which exposes data as assets (a markable entity with name and address), and uses standard HTTP methods to create, load, update and delete them.

In Golang, the Gin request library provides an interface for the RESTful APIs and the Webhook system. We use this library instead of the built-in http library because it offers much more functionalities to deal with GitHub Webhooks and a great integration with the library github. There are three basic methods for the API, which are listed below.

• GET. This method allows the client to load data (asset) from the server.

• POST. This method allows the client to create data (asset) on the server.

• ANY. This method registers a route that matches all the HTTP methods, including GET and POST.

In our code, we use the GET method for users to get the current status of our queuing system, use the POST method for GitHub to deliver the Webhook to us, and ANY to act as an interface of OAuth2 authentication.

For the GET method, if we want to pass arguments into the callback function, we can use the Query attribute. After the request is sent, the two global variables Name and Grade in Go runtime environment will be assigned the values specified in the request. For example, if a request ../gradetest?name=jack&&grade=100 is received by the server, the name and grade fields will be parsed by the Go runtime program and be stored by Name:=c.Query("name")and Grade:=c.Query("grade").

## 2.4 The Producer-Consumer Model

In our work, there are many opportunities for parallelism, as another task may come before the current one is executed, and modern CPUs are probably multi-core. When tasks come within the parallel threshold, we assign a consumer to the task at once, and the consumer is thus no longer available until the process ends. When more tasks than the threshold come, we put them into the queue (and that's why our framework is essentially a queuing system).

Our queuing system is basically an implementation of the producer-consumer queuing model. When a task comes, it is put into the channel, or we call it producer enqueues a task. When a thread is available, it takes the tasks and marks itself as unavailable, or we call it consumer dequeue a task, which makes the channel size decrement by 1. When all the consumers are not available, the queue will not be empty, and tasks pile up in it. When the queue is full, the program ends with an exception.

Related problems concerning locks immediately show up in this case. In this part, we mainly talk about two patterns: the blocking channel and the blocks. As Golang is a server/web programming language, blocking is inevitable. If we want the goroutines to report their process, we need to use blocking strategies. Golang provides such good tools to make this easy: channel.

Here we only discuss the synchronizing channel without buffer. To announce such a channel, use the code below to create a channel containing integers: ch := make(chan int). To use this channel for delivering data or blocking, we need to define the input and output port of this channel, as shown below.

```
func main(){
  go func() {
     // some web IO here
     ch <- a
  }
  x := <- ch
  print("successful!")
}
```

Because there is no buffer, the input port, and the output port will be blocked if there is no input for the channel. In this case, the program guarantees that the print command will be executed only after the web IO operations are done, i.e., after a is delivered into the channel ch. We used this approach in our framework to make all the consumers wait for a task to enter the queue, and execute it when the consumer is available.

There are some cases where we need to share resources among different threads, and without a lock, it's rather dangerous. In this case, we use the mutex or RWmutex in Golang to avoid conflict. When we use the mutex lock, when a thread obtains the lock, all other threads must wait for this thread to release the lock to obtain their lock, as there is only one lock. When we use the RWmutex, when a read lock is obtained, we can still obtain other read locks, but we can't obtain a write lock; when a write lock is obtained, we can't obtain either a read lock or a write lock.

## 3. Project Architecture

In this part we discuss about the technical details in our extension, including the overall architecture we developed, the programming tricks, the bugs we met, and also the issues we resolved. The program contains a large variety of Golang knowledge features like go routine and Webhook programming. The code is available at https://github.com/BiEchi/GoAutoBash.

### 3.1 The Main Program

The main program in this project is called main.go, and is the entry program for this project. Looking into the code, you will find that it basically does two things: starting the queuing system and listening to the Internet and VCS.

### 3.2 Start the Queueing Process Parallelly

In this part, we begin communications with the course GitHub server. Firstly, we make a new channel (queue) of tasks. Then we launch the specified number of consumers, each iterating through all the tasks in the queue. As we mentioned in Section 2.4, the channel is blocking, so each consumer(go routine) waits for a task to enter the channel. When a task enters the queue, the consumer begins to execute automatically. After this iteration, the consumer comes back to the for statement and executes the next task which is the head of the queue if it exists. If the queue is empty, the consumer starves and waits for another task to enqueue. When several consumers are starving and waiting for a task to enter the queue, which consumer gets the task does not matter. The illustration is in Figure 3.

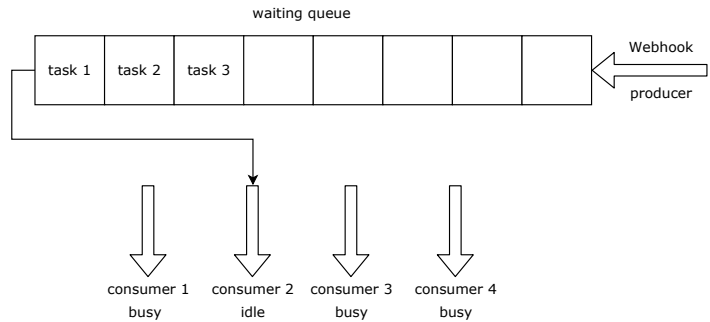

Figure 3. The producer-consumer model of parallelism in our work. The Webhook, as a producer, assigns tasks to the waiting queue (channel). The threads, as consumers, takes the tasks from the waiting queue for parallel execution.

Secondly, to avoid race conditions, we utilize mutex for synchronization. We divide the consumer workflow into three phases: reading, executing (testing subroutine), and writing. When in the reading phase, we request the read lock, and we save the shared variables to a local copy. In the testing subroutine, we utilize the local copy for functionalities. In the writing phase, we upgrade the shared variables by firstly obtaining the write lock. By this approach we avoid preempting the whole shared variables throughout the workflow of a consumer and improves performance greatly. The graph is shown in Figure 4.

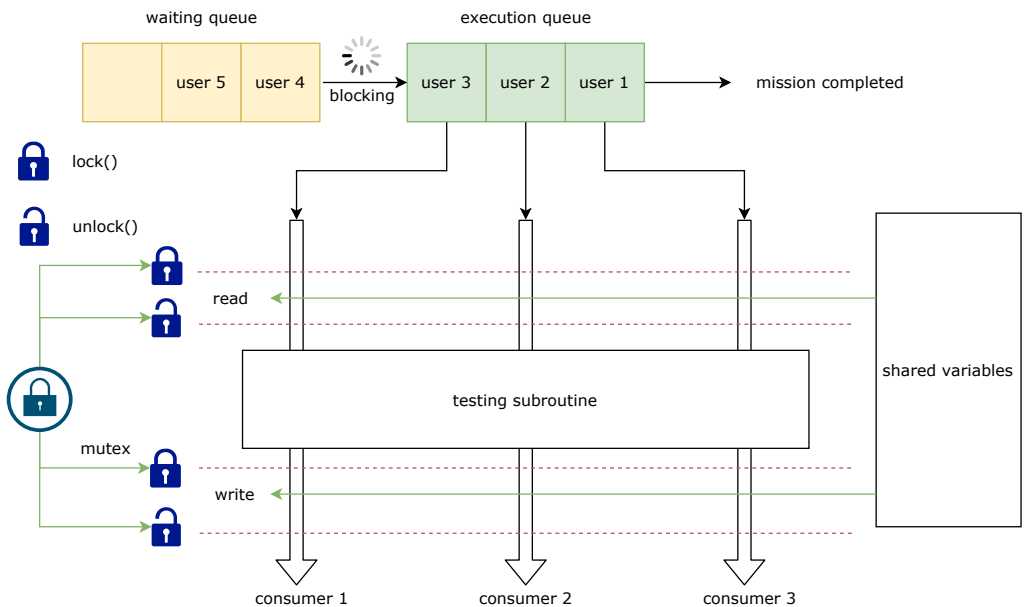

Figure 4. The architecture of the locking strategy. One read/write mutex is used to solve the potential danger when parallelly changing the values in the shared variables.

Note that compared to the grading subroutine, other logics are not time-consuming and thus the locking will not block a thread so much time that excesses the time of the grading subroutine. In this case, our queue system is efficient.

### 3.3 Listen to Internet and VCS

In this part, the program keeps running and listening to the browser. The API list is shown in Figure 5, where each function is a handler for the request. When the user sends a GET request, the green callback functions will be executed. When GitHub sends a Webhook POST request, the red callback function will be executed. The yellow callback function will be executed in either case, but only useful when communicating with the VCS OAuth2 server, which will be explained in more detail below.

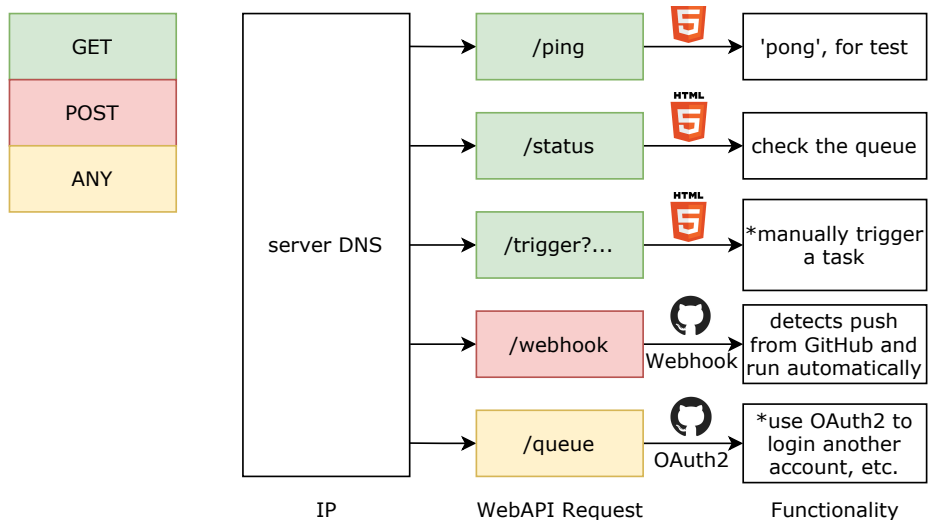

Figure 5. The graphical illustration of the listen list. The requests with "*" means it's can be extended but not implemented in our framework.

The ping request is a test request for checking whether the server is running the auto grader. If the browser shows "pong", then the service is functioning well. If it returns no data or wrong status, then the service is not running or malfunctioning.

The status request is a request that can be used by anyone who knows the domain of this server. Users can use this request to observe how many tasks are currently in the waiting queue and what's the last one being executed.

The trigger request is a manual trigger that tells the server to run a specific assignment for someone. The user needs to add some arguments to specify what to do exactly, such as ../trigger/?a=1&&b=2. However, this is not recommended as it's under risk of penetration attacks.

The webhook request is the core functionality in this system. After the student pushes his repository with his code assignment, the server detects the change in the code, and runs the subroutine to grade the latest version of the corresponding assignment automatically, which is equivalent to adding an assignment into the channel shown in Figure 4.

The queue request is only for private uses. With the queue request, a user can view his own testing process if it is currently under execution. Considering this, it's recommended to implement the functionality utilizing the OAuth2 approach that requires the user to log in before they can check. After logging in, students will have the token in their browser and thus they're allowed to process the request. In this way, a user cannot get others' information unless they have their GitHub account and password.

## 4. Application: KLC-3 Grading System

After explaining the framework we now introduce an application of our framework: the KLC-3 Grading system developed by Wenqing Luo [6] for a coding course at the University of Illinois at Urbana-Champaign. The source code of the queuing system is inherited from our framework and a copy is available at https://github.com/BiEchi/GoAutoGrader under the UIUC license. In the course ECE220 of UIUC (Computer Systems & Programming), students are required to write LC-3 (a type of assembly language) and C code for coding assignments. Giving immediate feedback to such assignments manually is impossible due to a large number of requests, so scripts were developed to facilitate the grading process. The script is designed so that, we feed both the student code and gold code (standard solution code) with the same input and expect the same output from both versions. A graphical illustration is shown below in Figure 6.

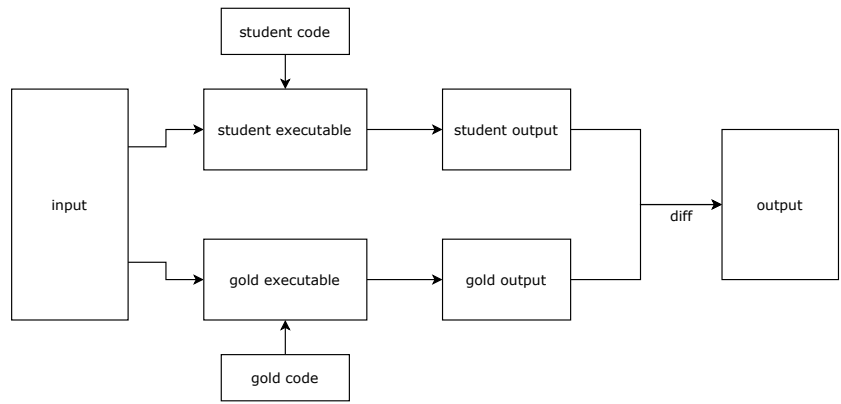

Figure 6. The graphical illustration of the script method for handling students' code submissions. Input: Student code. Output: Difference between results of the two codeworks.

A more concrete graphical illustration is shown in Figure 7. With the input of the student code, gold code, and a constraint of assignment specification, the tool is able to call the KLEE backend to analyze the possible issues and bugs in the student code and generate a report for the student code. Note that all the students have an account for the GitHub Enterprise server using LDAP authentication.

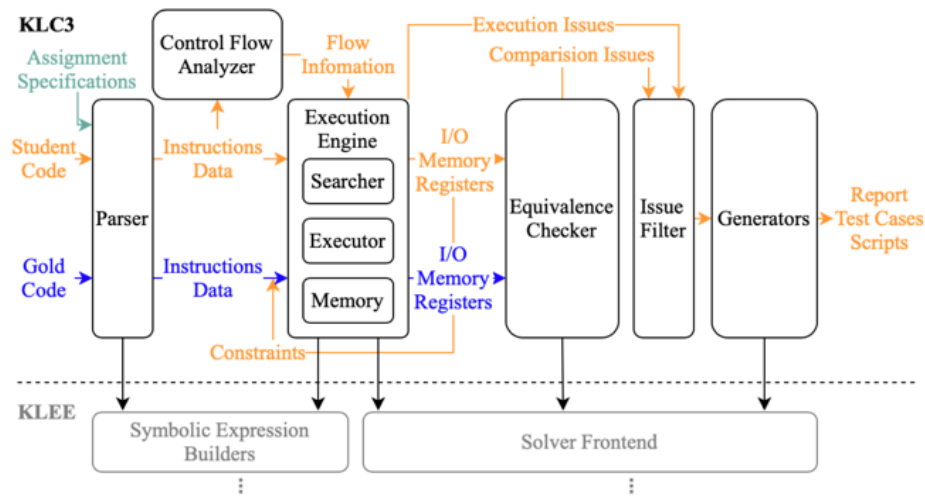

Figure 7. The graphical illustration of KLC3 [6]. KLC3 is built upon KLEE infrastructure, and changed the frontend architecture to facilitate KLEE to work on LC-3 assembly code.

Moreover, the KLC-3 project needs the testing server to synchronize its assets of students' history code and symbolic database with another repository on GitHub, so every time after releasing the write lock the consumer needs to push the contents to the VCS as well, which is illustrated in Figure 8.

After the students push the code repo to GitHub Enterprise, it sends a Webhook to the course server and triggers a grading task. The server then synchronizes the symbolic database by pulling the symbolic data from the private GitHub repo and use the code on GitHub enterprise and symbolic data to run KLC-3 grading. After the grading script is done, the server pushes the new symbolic data back to the private GitHub repo, and pushes the report to the GitHub Enterprise repo, which can be accessed by a student.

As we see, the KLC-3 grading script is performance-demanding as each task takes about 4-5 minutes [6], and when approaching the deadline of a coding assignment, the transaction number increases rapidly.

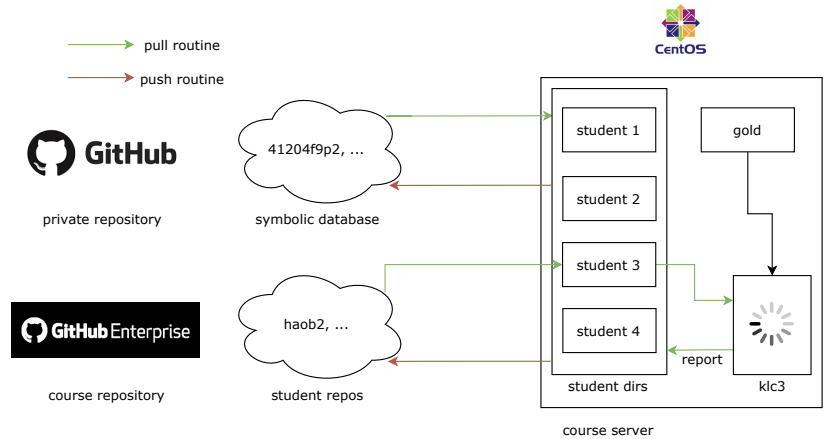

Figure 8. The graphical illustration of KLC3 [6]. KLC3 is built upon KLEE infrastructure, and changed the frontend architecture to facilitate KLEE to work on LC-3 assembly code

## 5. Conclusion

In this paper, we introduced the testing framework called GoAutoBash that detects the specified behaviors on VCS and processes the testing scripts automatically after receiving the Webhook signal from VCS. We looked into the reason we use Webhook, Golang and the Gin request library. We then discussed the main architecture of our project: parallelly running the assignments in the queue system of the server, and keep running for listening to certain tasks dispatched. We also looked into a specific implementation of our framework: the KLC-3 grading system, from which we know why we need such a queuing system for performance.

## Acknowledgement

Thanks to Wenqing Luo, a great developer and teaching assistant of ECE220, who provided me great help in understanding his code in *GoAutoGrader*. Professor Steven S. Lumetta offered great help in my communications with Luo too. I also need to thank Nuohan Yang, who gives me so much support in my spirit to accomplish my work.